# A generalization of the inhomogeneity measure for point distributions to the case of finite size objects

Ryszard Piasecki[*]

*Institute of Chemistry, University of Opole, Oleska 48, 45-052 Opole, Poland*

**Abstract**

The statistical measure of spatial inhomogeneity for $n$ points placed in $\chi$ cells each of size $k \times k$ is generalized to incorporate finite size objects like black pixels for binary patterns of size $L \times L$. As a function of length scale $k$, the measure is modified in such a way that it relates to the smallest realizable value for each considered scale. To overcome the limitation of pattern partitions to scales with $k$ being integer divisors of $L$ we use a sliding cell-sampling approach. For given patterns, particularly in the case of clusters polydispersed in size, the comparison between the statistical measure and the entropic one reveals differences in detection of the first peak while at other scales they well correlate. The universality of the two measures allows both a hidden periodicity traces and attributes of planar quasi-crystals to be explored.

*PACS:* 05.90.+m

*Keywords:* Spatial inhomogeneity; Multiscale analysis; Binary patterns; Planar quasi-crystals

## 1. Introduction

With the help of composition variance analysis the early method of quantitative homogeneity characterization of binary grain mixture by so-called mixing index was worked out by Lacey [1]. The author proposed also a definition of the intensity of segregation applying variance analysis of the numerical fraction of a one of two components in a mixture [2]; see also [3] for a review. Further, the corresponding multiscale sampling image analysis was developed for both the intensity and the scale of segregation in a binary particle mixture [4]. Additionally, the expected and experimental variations in the mixing index at different sampling sizes and for different times in a mixing process were presented. As concerns more practical aspects of the image analysis of the mixture sections, in the latter paper the grains were reduced to a point by homotoping marking. Thus the spatial homogeneity characterization was related to a *point set* (pattern of points)

---
[*]Fax: +48 77 4527101.

 *E-mail address:* piaser@uni.opole.pl (R. Piasecki).

representing two-phase grain mixture. To the author knowledge, this approach is the ubiquitous one.

Recently, originating from physics the entropic measure $S_\Delta$ of spatial inhomogeneity for patterns of point objects (PO) [5] has been generalized to the case of finite size objects (FSO) [6]. For binary patterns FSO are represented by black pixels. In fact, we generalize the method for systems of non-interacting point-like particles to systems of identical objects 'interacting' with each other through mutual exclusion (via hard-core repulsion). In contrast to statistical analysis of irregular point patterns or point process statistics [7–11], no *a priori* assumptions are made about distribution of objects. The generalized measure $S_\Delta$(FSO) can be used in searching for more accurate correlations between macroscopic properties of spatially inhomogeneous real or model media and their microstructure features [12], for *any concentration* of FSO.

This relatively short paper does not attempt to discuss all aspects of the spatial inhomogeneity analysis of binary patterns. The main aim is to show that one of statistical measures $h$ [13] developed for PO-patterns can be also generalized to FSO-case $h_A$, see a brief information in Ref. [6], Appendix B. Further its modification $h_\Delta$(FSO) allows us to make a reasonable comparison with the entropic counterpart $S_\Delta$(FSO) mentioned above. In connexion with Ref. [6], the present work is motivated by its complementary function to the paper [5], where the PO-versions of the two measures have been already compared.

The paper is organized as follows. In Section 2 we present the statistical measures $h$(PO) and $h_\Delta$(FSO) in a slightly changed notation in comparison with Ref. [6]. Also a sliding cell-sampling approach used in this paper is briefly sketched. Section 3 deals with the inhomogeneity analysis for a number of simulated or adapted patterns. The results for statistical measure $h_\Delta$(FSO) and entropic one $S_\Delta$(FSO) are compared. Finally, in Sec. 4 the concluding remarks are given.

## 2. Modification of the point measure

To analyse spatial homogeneity for a distribution of $n$ points a space is subdivided into $\chi$ cells of equal size $k \times k$ with the occupied numbers $n_i$ and then the sample variance is compared with its theoretical value [14]. Another approach for point patterns was given in detail in Refs. [13,15,16]. It compares the actual random variable $\mu(\text{PO}) \equiv \Sigma_{i=1}^{\chi}(n_i - n/\chi)^2$ with its expected value leading to the statistical measure of spatial inhomogeneity for patterns of point objects $h(k; \text{PO}) \equiv \mu/E(\mu) = \chi\mu/[n(\chi-1)]$. We underline that the latter formula is formally equivalent to that one obtained from a different point of view by Zwicky [14].

However, when the pixels are counted correctly as FSO, the calculation of the expected value of the random variable $\mu_A$(FSO) needs a modification. Certain configurations possible for points, e.g., all points placed at the same cell, cannot be realized for pixels due to their finite size. For the occupation numbers $n_i(k)$ of black pixels besides the standard constraint $n_1 + n_2 + ... + n_\chi = n$ the additional one, $n_i \le k^2$, should be fulfilled for every $i$-th cell. In this case the multicomponent generalization of the hypergeometric distribution give us (the alternative but simple author's way of obtaining of the $E(\mu_A)$ is briefly presented in Appendix)

$$h_A(k;\text{FSO}) \equiv \mu_A/E(\mu_A) = c(k,n)\,\frac{\chi\mu_A}{n(\chi-1)} \ , \tag{1}$$

and its final modification, $h_\Delta \equiv (\mu_A - \mu_{A\,\text{min}})/E(\mu_A) = h_A - h_{A\,\text{min}}$, that evaluates the deviation of the actual configuration from the possible maximally uniform configuration for underlying scale $k$ reads

$$h_\Delta(k;\text{FSO}) = c(k,n)\,\frac{\chi}{n(\chi-1)}\left[\mu_A - r_0\,(\chi-r_0)/\chi\right] \ . \tag{2}$$

In a slightly different form the above formulas have been briefly announced in Ref. [6], (see Appendix B, cf. Eq. (B2)), where unfortunately the last term is included with a misprinted sign; it should be read as $2n_0 + 1$. For a given length scale $k$ the values of the FSO-factor $c(k,n) \equiv (L^2-1)/(L^2-n)$ with $L^2 = \chi k^2$ ranges from 1 (for $n=1$) to $L^2-1$ (for $n=L^2-1$). The reference measure $h_{A\,\text{min}} \equiv \mu_{A\,\text{min}}/E(\mu_A)$ corresponds to the lowest *realizable* value of $h_A$ at a given length scale $k$. One can find it by fixing $n_0$ and $n_0 + 1$ black pixels in $\chi - r_0$ and $r_0$ cells, where $r_0 = n \bmod \chi$ and $n_0 = (n-r_0)/\chi$. To obtain the corresponding modified PO-measure $h_\Delta(k;\text{PO})$ is enough make the replacements $\mu_A \to \mu$ and $c(k,n) \to 1$.

Now $h_\Delta(k=1;\text{FSO}) \equiv 0$ and $h_\Delta(k=1;\text{PO}) \equiv 0$ since for $k=1$ we have $r_0 = n$, thus $\mu_A = \mu_{A\,\text{min}}$ and $\mu = \mu_{\text{min}}$. For the second boundary scale $k = L$ we have $r_0 = 0$. Taking a limit $\chi \to 1$ one can obtain $h_\Delta(k=L;\text{FSO}) \to 0$ and $h_\Delta(k=L;\text{PO}) \to 0$. For a given scale $1 < k < L$ all possible values of the statistical FSO-measure range from 0 for the most uniform configuration of $n$ pixels to $c(k,n)\,n$ indicating the most inhomogeneous one. A peculiar case appears when $\mu_A = E(\mu_A)$. Then, according to formal definition, such an arrangement of black pixels can be treated as perfectly random configuration

$$h_\Delta(k;\text{FSO-random}) = \begin{cases} 1 - c(k,n)\,r_0\,(\chi-r_0)/n(\chi-1) & \text{for } 1<k<L \\ 0 & \text{for } k=1,L \end{cases} \tag{3}$$

The knowledge of such a length scale that detects the perfectly random configuration may have some meaning for discussion on possible connection between randomness and pattern complexity, see for instance Refs. [23,24,25]. Again, to obtain the corresponding PO-case the replacements $\mu_A \to \mu$ and $c(k,n) \to 1$ should be made.

Notice that if the point measure also applies to a binary pattern we have $\mu \equiv \mu_A$ and the two statistical measures are quantitatively distinguishable just by the FSO-factor $c(k,n)$ expressed here as a function of concentration $\varphi = n/L^2$ of black pixels

$$h_\Delta(k;\text{FSO}) = \frac{(1 - \varphi/n)}{(1 - \varphi)}\,h_\Delta(k;\text{PO}) \ . \tag{4}$$

Generally, for the same binary pattern, despite of the *quantitative* differences between the exact measure $h_{\Delta}(k;$ FSO) and the approximated $h_{\Delta}(k;$ PO) the usage of the latter one is expected to be roughly *qualitatively* correct. However, within the standard partitioning for the both statistical measures the limitation to the scales for which $k$ is an integer divisor of $L$ still remains.

This weakness has been recently overcome by introducing entropic average measure (per cell) of spatial inhomogeneity $S_{\Delta}(k) \equiv (S_{\max} - S)/\chi$, see Refs. [5,6]. Here $S_{\max}$ and $S$ describe respectively the highest possible configuration entropy and the real one for a given pattern; see also Refs. [17–22] for other applications and extensions. It was possible due to specific property of the entropic measure $S_{\Delta}$ that conserves its value when an initial pattern $L \times L$ is periodically repeated $m$-times. This is not the case for the present generalization.

Therefore, instead of the standard partitioning we use a *sliding* cell-sampling (SCS) approach, see for instance Refs. [26–28]. Under the name "gliding-box" the method was used to found the fluctuations of the mass distribution function to define the lacunarity of a fractal set [26]. In turn, by means of explained in detail the "sliding window" concept, the study of human EEG-signals was undertaken by recourse to a wavelet based multiresolution analysis employing time dependent Shannon's and Tsallis's entropies [27]. Recently, an algorithm to estimate the Hurst exponent of high-dimensional fractals based on a generalized high-dimensional variance around a moving average over a one-dimensional (1D) "sliding window" called a "sub-array" in 2D case has been tested for rough surfaces [28].

Within the SCS approach, for a sliding factor $1 \leq z \leq k$ the number of $\chi_a \equiv [(L-k)/z + 1]^2$ cells is sampled, each of size $k \times k$, provided $(L-k) \bmod z = 0$. Here we perform calculations for $z = 1$ that gives the maximal overlapping of the sampled cells. Actually, the statistical measure given by Eq. (2) and the entropic one, see Ref. [6] (cf. Eq. (4)), we apply to auxiliary patterns $L_a \times L_a$, where $L_a \equiv [(L-k)/z + 1] \, k$. For every length scale $k$ the auxiliary pattern is composed of the sampled cells placed in a non-overlapping manner. Such a pattern clearly reproduces the general structure of the initial one, see Fig. 1b in Section 3. Therefore, we may treat it as a representative pattern for investigated image. Now, for a given scale $k$ the sum of occupation numbers $n_i(k)$ depends also on the sliding factor, i.e. $n = n(k; z)$ until $z < k$. From this viewpoint, when the sliding factor $z = k$ there is no overlapping and we have $L_a = L$. In this case the SCS approach is exactly the same as the standard partitioning. For other allowed $z$-values, i.e. for $1 < z < k$, the enriched multiscale inhomogeneity analysis seems to be an interesting topic for the future study. This kind of filtering of the whole set $\{n_i(k)\}$ of cell occupation numbers can provide a tool for searching, for example, traces of statistical self-similarity hidden in a sub-domain of initial pattern.

Two more remarks are in order. The SCS approach involves a certain averaging process since some black pixels are common for the neighbouring positions of the sliding cell. For this reason it provides rather smooth but still useful characteristics for an arrangement of objects over entire range of the length scales. We should also remember that for the scales close to $L$ the number of sampled cells $\chi_a$ may be not large enough in comparison with the number $N_a$ suitable for a good cell statistics. Nevertheless, for a given number $N_a \leq \chi_a$ one can always use the simple inequality

$$k \; \leq \; L \, - \, z\left(\sqrt{N_a} \, - \, 1\right), \tag{5}$$

to estimate: (*i*) the limit length scale $k$ for a given $L$, or (*ii*) the minimal size $L$ of a pattern for assumed $x \equiv k/L$. For example, taking the sliding factor $z = 1$, $N_a = 1000$ we obtain a practical rule for the limit scale $k = L - 31$, leading to $k_a = 329$, 550 and 670 when $L = 360$, 581, 701. On the other hand, for the same $N_a$ and various $x = 0.9$, 0.92, 0.94, 0.96 we need for statistical meaning a pattern of size $L = 307$, 383, 511 and 766 at least.

## 3. Examples

We remind that our measures as a function of length scales $k$ are discrete ones and the continuous lines in the figures should be treated as a guide for eyes. For testing purposes of the SCS approach we start with prepared by a cellular automata 83×83 pattern shown in the inset in Fig. 1a. Notice, that there is no periodicity for the particle configuration. Secondly, the length of pattern side is chosen to be a prime number. Fig. 1a shows the two well-correlated curves, solid for $h_\Delta(k; FSO)$ and dropped one for $h_\Delta(k; PO)$. According to Eq. (4) we have $h_\Delta(k; FSO) > h_\Delta(k; PO)$ except the two boundary scales, $k = 1$ and $k = L$, for which there is an equality. Additionally, for $1 < k < L$ the horizontal line close to 1 denotes the value of the measure described by Eq. (3) for configuration of black pixels distributed completely at random. The corresponding PO-line is not shown since it is not distinguishable with a naked eye from the FSO-case. The interesting behaviour one can find for the middle three ranges of scales around the peaks at $k = 32$, 44 and 54. The values of FSO-measure deviate from the limit of perfect randomness toward to the higher values in contrast to the behaviour of PO-measure. This means that at these intervals the spatial inhomogeneity is greater than occurs for a perfectly random configuration of pixels while PO-measure leads to a reverse conclusion. Although it is limited to several scales only, this is the first *qualitative* inconsistency found in a usage of PO-measure to the FSO-configuration.

Concerning the first maximum, clearly higher than the others, in the $h_\Delta(k; FSO)$ and $h_\Delta(k; PO)$ curves it is reasonable to interpret it as an indicator of small cluster formation. In Ref. [29] such interpretation was argued for another measure of morphological features named normalized information entropy $H' \equiv H - H_r$. The expected information entropy $H_r$ of a presupposed random configuration is subtracted from the information entropy $H$ calculated for a real particle distribution. In our case, the pixels clustering at $k_{cluster} = 6$ causes large values of both measures by increasing the number of cells almost filled and the number of nearly empty cells in comparison with those expected for a uniform distribution. In Fig. 1b the auxiliary pattern of size 468×468 obtained for $k = k_{cluster}$ illustrates this situation. Also the additional maximums in Fig. 1a occur at scales at which the pixels distribution deviates distinctly from a uniform one. One can say that processes of grouping of clusters are prevalent at these scales.

In turn, the minimums correspond to ordering of pixels. They indicate that the pixels distribution at these scales is relatively close to a uniform one. Periodicity of a pattern can be identified if *all* minimums in the $h_\Delta(k; \text{FSO})$ curve (also in $h_\Delta(k; \text{PO})$ one) are equidistant from each other. For both curves minimums appear at the same scales, namely $k = 13, 24, 39, 49, 60$ and $72$. Taking into account the number $N_a = 1000$ as suitable for a good cell statistics the range of scales $1 \le k \le 52$ results from Eq. (5) for statistical meaning. It may be noticed in Fig. 1a that within this range the inter-distances between the minimums differ from each other. This means the lack of a periodicity of the analysed configuration. It is interesting that for considerably weakened statistical meaning condition, e.g. with the lower number $N_a = 500$ leading to a larger range of allowed scales $1 \le k \le 60$, barely two inter-distances have the same length equal to 11 pixels. However, the above statement about lack of a periodicity still remains true.

From here we focus on the FSO-measures, the statistical $h_\Delta(k; \text{FSO})$ and mentioned above the entropic $S_\Delta(k; \text{FSO})$. The latter one has been already applied with the same interpretation of its maximums and minimums [6,19−22]. However, it should be stressed that the standard partitioning procedure has been then used exclusively. Now, we pay special attention on the sensitivity of compared measures for periodicity feature. Let us consider four computer generated patterns, each of size 360×360 in pixels. The patterns (A) and (B) are locally random while the (C) and (D) are globally regular. The all collection is shown in Fig. 2. To create the first three patterns the area was subdivided into 30×30 lattice cells. For pattern (A) a group of 120 black pixels is randomly spread around the centre each of the lattice cells. The statistical periodicity of the pattern is evident. More spatially inhomogeneous pattern (B) consists of compact clusters each of 120 black pixels too. The clusters are randomly placed, one per lattice cell. Now, statistical periodicity is hardly seen and may be called the hidden one. In turn the regular pattern (C) includes again the same compact clusters but each of them occupies the centre of lattice cell. In this way we obtain exactly periodic pattern of structure a square lattice with the lattice constant $k_{\text{const}} = 30$ pixels. For regular pattern (D) of structure a triangular lattice again with the same clusters and identical lattice constant there is no square basic cell. Instead, with a sufficient accuracy for our purposes this periodic pattern one can obtain using as a building block a rectangle of size 30×52 with a full compact cluster and two matched cluster parts inside of it. This basic cell is shown in the inset in Fig. 3c.

In Fig. 3a the corresponding values of both measures: $h_\Delta(\text{B}) > h_\Delta(\text{A})$, thick solid curves and $S_\Delta(\text{B}) > S_\Delta(\text{A})$, thin solid curves, quantitatively confirm the different degrees of spatial inhomogeneity for the patterns. Notice also, that for (A) all minimums of the both measures are equally distant from each other. The inter-distance equals to 30 pixels and clearly coincides with the statistical periodicity of the configuration. On the other hand, the hidden statistical periodicity of (B) is not visible at first sight due to the local randomness of positions of the compact clusters. Both curves $h_\Delta(\text{B})$ and $S_\Delta(\text{B})$ oscillate quite regularly at large length scales clearly suggesting occurrence of a periodicity.

However, the strongly asymmetric first peak makes the situation unclear. Nevertheless, such a periodicity can be also extracted in a simple way that was described in Ref. [30]. Namely, one can decompose $h_\Delta(B)$ and $S_\Delta(B)$ into a superposition of a few contribution functions to obtain by the fitting procedure the sequential peak positions $k_{fit}$ except the position of the first asymmetric peak. Then, from a linear best fit of the peak positions $k_{fit} = (l + 0.5) k_{const}$, where $l$ denotes odd numbers one can deduce a lattice constant $k_{const}$ linked with a periodicity. Thus its value should agree with the inter-distance between minimums at large scales. Recently, in this way the first *quantitative* evidence for the presence of a quasi-regular pattern in the photospheric intensity fields has been obtained [30].

We should also point out that in Fig. 3a for pattern (A) the heights of the first maximums of the both measures, both at $k = 15$, are only slightly higher than the additional ones. As expected, the contrast results appear for pattern (B). Now the compact cluster presence is clearly marked around two scales, $k_{cluster}(h_\Delta) = 14$ and $k_{cluster}(S_\Delta) = 17$, uncovering in this way subtle differences in detecting of compact clusters by the two FSO-measures. It is worth to mention that some kind of a complementary behaviour of $h(PO)$ and $S_\Delta(PO)$ was found in Ref. [5]. There, for a fixed number of points placed at four cells among all representative configurations a few pairs of the configurations distinguished by $S_\Delta(k; PO)$ but not by $h(k; PO)$ and inversely, have been revealed.

In Fig. 3b two thick solid curves of the measure $h_\Delta(k; FSO)$ and in Fig. 3c two thin solid lines of the measure $S_\Delta(k; FSO)$ are calculated for the pair of the regular patterns (C) and (D) but of different lattice structure. The both measures show the similarities in the shapes, positions and values of the first maximum within range of scales 1÷30 pixels since the same compact clusters are used in computer simulation for these patterns. Within this range the influence of nearest neighbourhood of different symmetry is weak. However, for the other scales the oscillating behaviour of both measure curves in Figs. 3b, c, especially regarding the positions and values of the peaks, shows meaningful dissimilarities coming from the different structures of both patterns. However, the periodicities are still clearly seen in spite of a square cell-sampling that is not optimal for the triangular pattern (D). The lower spatial inhomogeneity of (D) in comparison with (C) is also confirmed by the corresponding values of the measure curves. Notice that the very similar behaviour of the normalized information entropy *H'* has been recently discussed for 200×200 testing patterns in Ref. [3], (cf. Figs. A.1, A.2).

Now, we would like to consider *aperiodic* two-dimensional patterns that are totally different from those already discussed and represent a quasi-crystal planar structure. Some authors put [31] an interesting question: how can we characterize the level of 'disorder' (or the degree of inhomogeneity) in well ordered aperiodic patterns, which are to be placed between the ideal crystals and amorphous media? At present, for the above-considered two measures we use a square cell-sampling approach. Supposedly, the more appropriate would be a sampling cell of a higher rotational symmetry. This kind of improved approach with a hexagonal sampling cell but for the normalized information entropy *H'* has been applied to investigation of photospheric structure in Ref. [32].

Fig. 4a depicts a sub-domain of size 701×701 adapted from Ref. [33] (cf. Fig. 2), where a two-dimensional cut-and-project set with 10-fold rotational symmetry was discussed. The binary pattern consists of the compact clusters similar in size to the already simulated ones. It represents a quasi-crystal planar structure that is free from a periodicity. However, a long-range positional order in some direction is still possible. In Fig. 4b the corresponding curves of the two measures, thick and thin solid one for $h_\Delta(k; \text{FSO})$ and $S_\Delta(k; \text{FSO})$, distinctly confirm the lack of periodicity. Notice, that the positions of minimums are more irregular than in Fig. 1a with 4-fold rotational symmetry. For both curves the first asymmetric peak has a hidden structure of two close located peaks. In the inset we compare the measure $h_\Delta(k; \text{FSO})$, thick solid curve, for initial 701×701 pattern with its four thin solid curves calculated for disconnected 350×350 sub-domains of the main pattern. The similarity of the curves with very small changes in the locations of peaks and minimums is still preserved for scales from 1 to about 280. It is interesting that we can see some traces of rough mirror symmetry in shapes of peaks and their positions around scales $k = 337\text{-}338$. The similar feature can be also observed in the one of the next figures. Our conjecture is that it may have a connection with the inverse symmetry (even though rough) of aperiodic patterns.

Fig. 5a presents a second example of aperiodic pattern. This is 581×581 sub-domain adapted from the reversed (black↔white) diffraction pattern of the octagonal Ammann-Beenker tiling discussed in Ref. [31] (cf. Box 5). The diffraction spots have different area that is proportional to the intensity of the diffraction peak. Some small spots have been discarded, i.e. those with an intensity of less than 0.05% of the intensity of the central spot. The pattern shows fairly accurate inverse symmetry. Supporting our conjecture the two measure curves in Fig. 5b show again approximate mirror symmetry in shapes of peaks and their positions within the central interval of length scales around $k = 238$. Another interesting feature is a shift in position of the first peak for the two measures. Namely, we have $k_{\text{cluster}} = 19$ for $h_\Delta$ and $k_{\text{cluster}} = 39$ for $S_\Delta$. When we analyse pattern with clusters equal in size the shift of such kind is negligible. This observation reveals a quite different behaviour of $h_\Delta(k; \text{FSO})$ and $S_\Delta(k; \text{FSO})$ at scales around the first asymmetric peak. The effect is particularly strong for aperiodic patterns consisted of clusters with size dispersion. We suppose that the different mathematical structure of the discussed measures manifests in this way.

## 4. Conclusion

In this paper we generalize and modify further the statistical inhomogeneity measure developed for point distributions to the case of finite size objects, see Eq. (2). To control cell statistics for the sliding cell-sampling approach the simple condition ensuring the appropriate range of length scales on dependence of the size of a pattern for a suitable number of sampled cells is obtained, see Eq. (5). For the most of chosen patterns the statistical measure and corresponding the entropic one, see Ref. [6] (cf. Eq. (4)), well correlate. However, for aperiodic patterns of a quasi-crystal planar structure at the length scales around the first asymmetric peak in the measure curves, particularly in the case of clusters with size dispersion, the correlation is much weaker. The statistical measure relatively overestimates the role of smaller clusters while the entropic one is comparatively more sensitive for larger clusters. However, it should be stressed that both measures equally well detect even hidden statistical periodicity.

## Appendix

The formula for probability of appearance of configurational FSO-macrostate $(n_1,..., n_\chi)$ has been already mentioned in Ref. [6] (cf. Eq. (2)). Although it is a multicomponent generalization of the well-known hypergeometric distribution, here the $E(\mu_A)$ is given in an alternative but simple way. By a definition of the expected value with a bit of algebra one finds

$$E(\mu_A) = f(n) - n^2/\chi, \qquad (A1)$$

where the function $f(n)$ is described by the recurrent relation

$$f(n+1) = \alpha_{n+1}\left[\beta_{n+1} - f(n)\right], \qquad f(1) = 1, \qquad (A2)$$

with $\alpha_n \equiv n/(\chi k^2 - n + 1)$ and $\beta_n \equiv \chi k^2 + (k^2 - 1)(n - 1)$. Its form can be easily determined

$$f(n) = n[(\chi + n - 1)k^2 - n]/(\chi k^2 - 1) \qquad (A3)$$

and the expected value finally simplifies to

$$E(\mu_A) = \frac{(L^2 - n)}{(L^2 - 1)}\frac{n(\chi - 1)}{\chi}, \qquad (A4)$$

where $L^2 = \chi k^2$. If instead of the standard partitioning we use a sliding cell-sampling approach then we make the formal replacements, $\chi \rightarrow \chi_a \equiv [(L-k)/z + 1]^2$, $L \rightarrow L_a \equiv [(L-k)/z + 1]k$ and $n \rightarrow n(k; z)$, where for a given scale $k$ the sum of occupation numbers $n_i(k)$ depends also on the sliding factor $z$, i.e. $n = n(k; z)$ until $z < k$.

**Figure caption**

Fig. 1. (a) Comparison of inhomogeneity measures $h_\Delta(k;$ FSO), thick solid curve, and $h_\Delta(k;$ PO), dropped one, for 83×83 pattern with 4-fold rotational symmetry given in the inset. For both curves the first maximum at length scale $k_{cluster} = 6$ indicates for small cluster formation. Notice that around the middle peaks at scales $k = 32$, 44 and 54 these curves deviate from the level of perfect randomness, horizontal line, in a quantitatively different way. For both curves minimums appear at the same scales given by $k = 13$, 24, 39, 49, 60, 72, which are not equally distant from each other. This fact can be linked with the lack of a periodicity of the pattern. For the number $N_a = 1000$ suitable for a good cell statistics the range of scales allowed for statistical meaning is $1 \le k \le 52$, see Eq. (5). (b) The auxiliary pattern of size 468×468 created with the sampled but non-overlapped 6×6-cells. It reproduces well the general structure of the initial pattern at the scale corresponding to the first maximum.

Fig. 2. Computer generated 360×360 patterns of clusters with periodicities. Every cluster consists of 120 black pixels. Patterns (A) and (B) are locally random patterns with spread and compact clusters while (C) and (D) are regular ones of structure a square and triangular lattice. For patterns (A) and (B) we have statistical periodicity easily seen by a naked eye and the hidden one, respectively.

Fig. 3. (a) Inhomogeneity measure $h_\Delta(k;$ FSO), thick solid curves, compared with $S_\Delta(k;$ FSO), thin solid curves, for statistically periodic patterns (A) and (B). The inset shows a one of 30×30-cells of the pattern (A). (b) For patterns (C) and (D) of different regular structure their periodicity is also clearly revealed by the measure $h_\Delta(k;$ FSO), thick solid oscillating curves, confirming different lattice constants $k_{const}$. (c) For the same patterns the behaviour of the entropic measure $S_\Delta(k;$ FSO), thin solid curves, leads to the same conclusion. Additionally, in the inset is shown 30×52 basic cell for the pattern (D).

Fig. 4. (a) Adapted 701×701 sub-domain of quasi-crystal planar structure with 10-fold rotational symmetry discussed in Ref. [33] (cf. Fig.2). (b) The two well correlating measures $h_\Delta(k;$ FSO), thick solid curve, and $S_\Delta(k;$ FSO), thin solid curve, clearly confirm the lack of periodicity. In the inset the measure $h_\Delta(k;$ FSO) for initial 701×701 pattern is compared with its four counterparts, thin solid curves, calculated for disconnected 350×350 sub-domains of the main pattern. The similarity of the curves appears for a wide range of scales, from $k = 1$ to about $k = 280$.

Fig. 5. (a) Adapted 583×583 sub-domain of the reversed (black↔white) diffraction pattern for the octagonal Ammann-Beenker tiling taken from Ref. [31] (cf. Box 5). (b) The two corresponding measures $h_\Delta(k;$ FSO), thick solid curve, and $S_\Delta(k;$ FSO), thin solid curve, well correlate except the length scales around the first asymmetric peak. Both measure curves also show distinctly the lack of a periodicity of quasi-crystal planar structure.

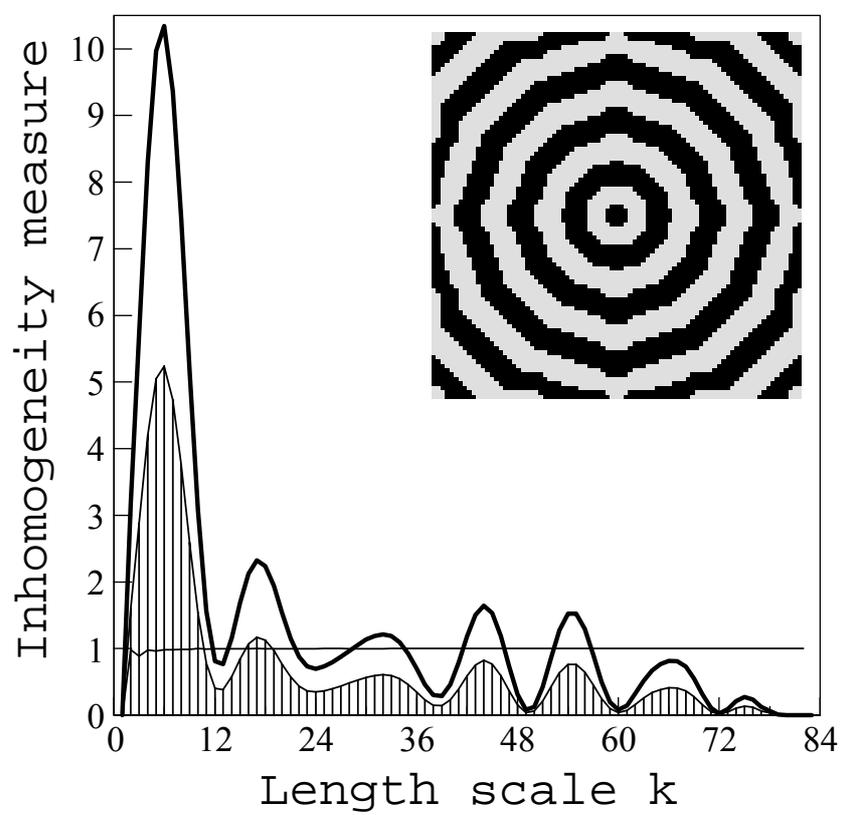

Fig. 1a

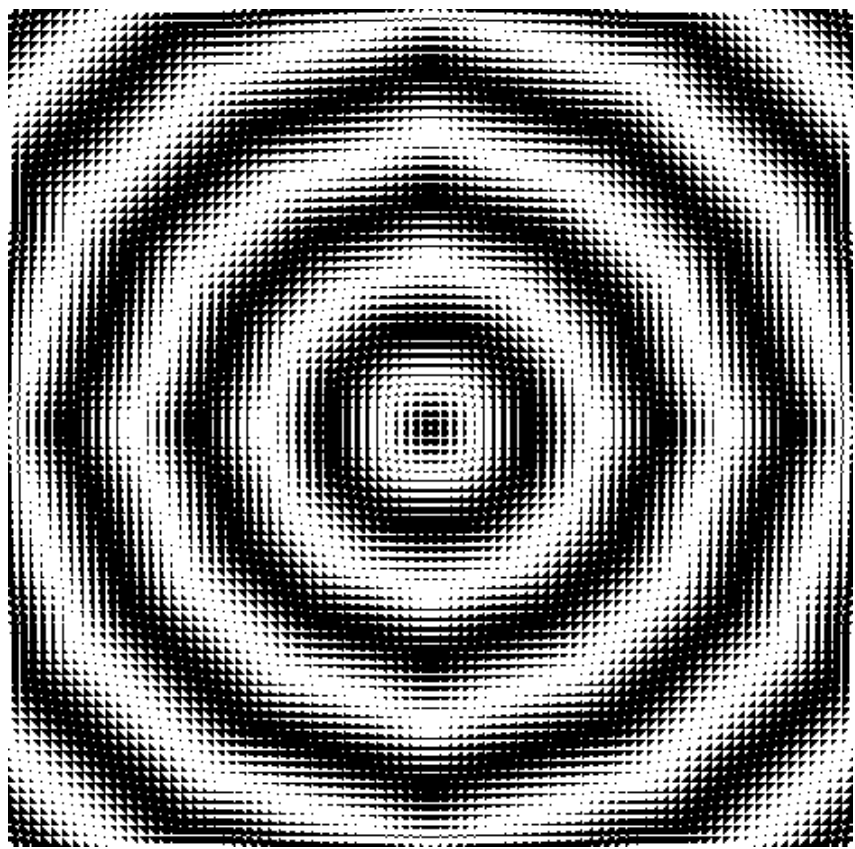

Fig. 1b

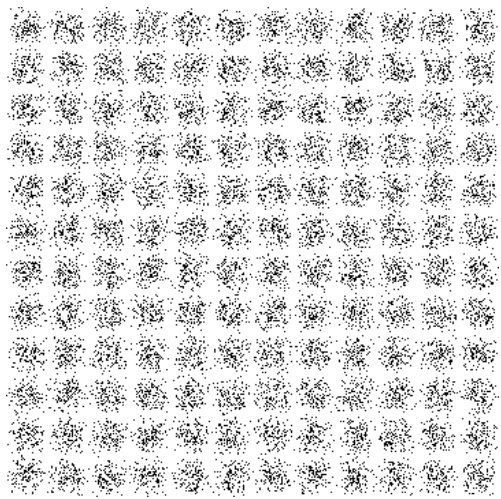

(A)

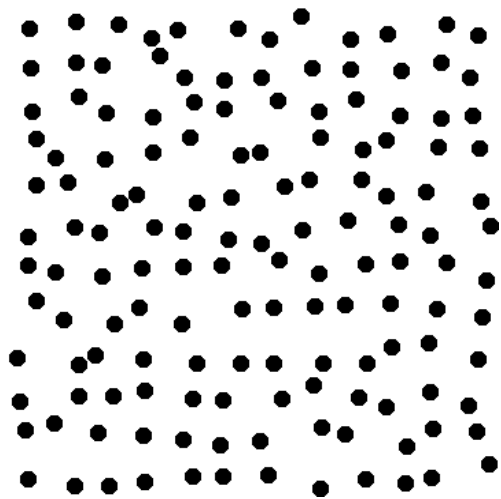

(B)

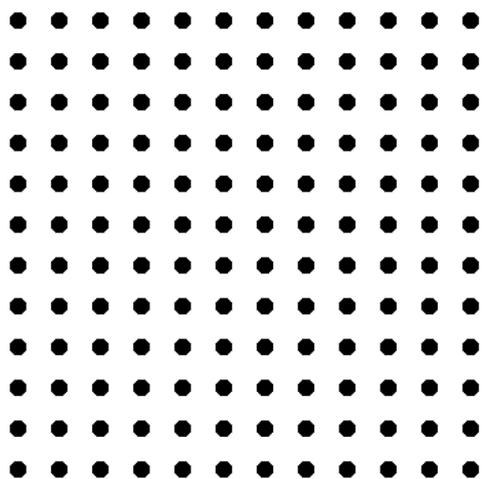

(C)

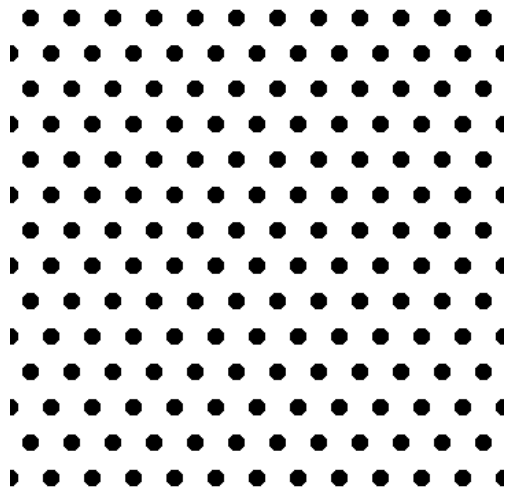

(D)

Fig. 2

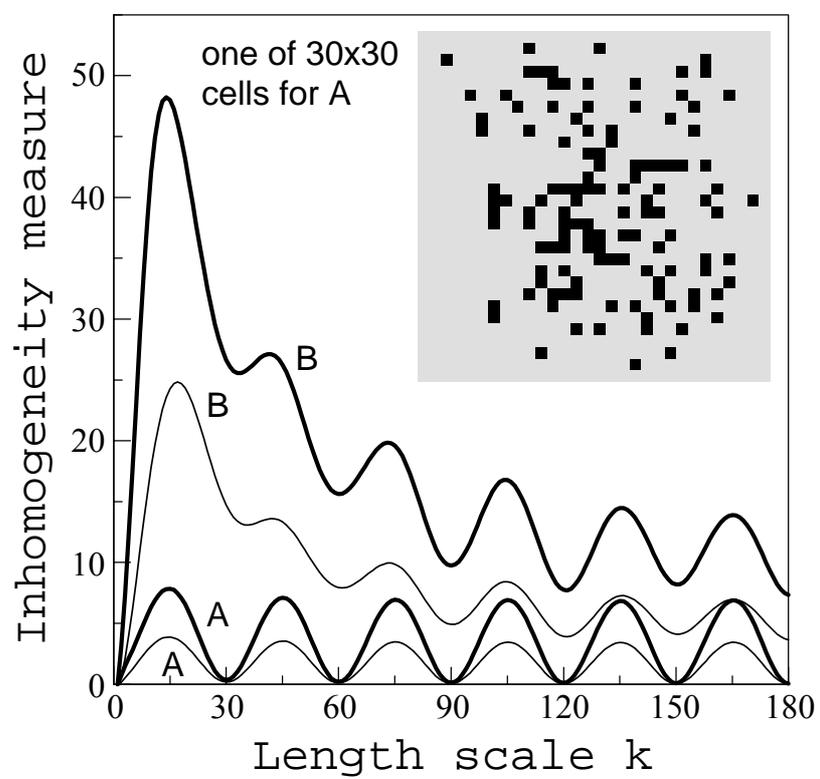

Fig. 3a

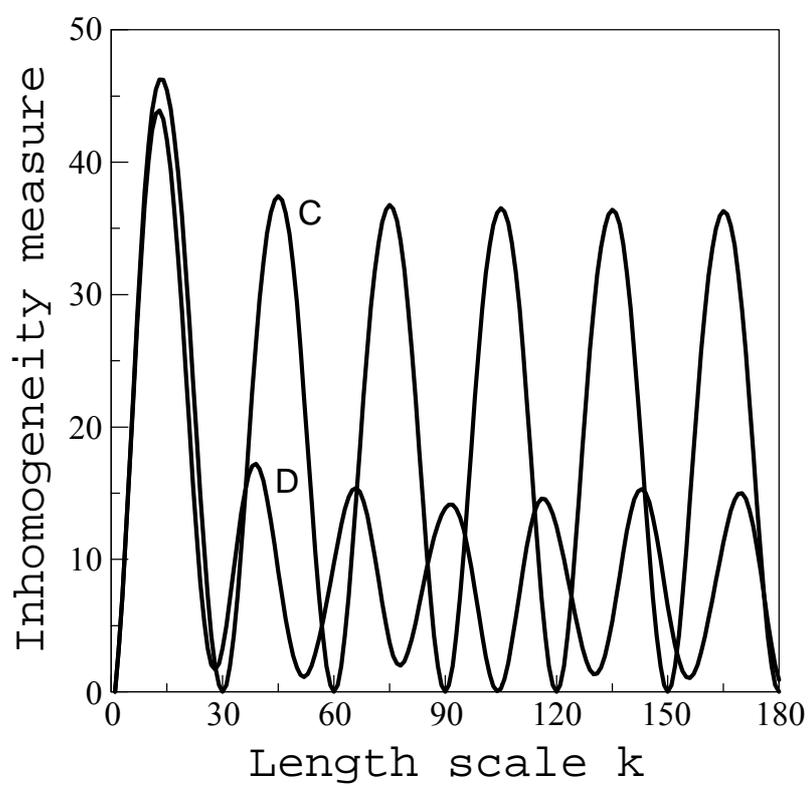

Fig. 3b

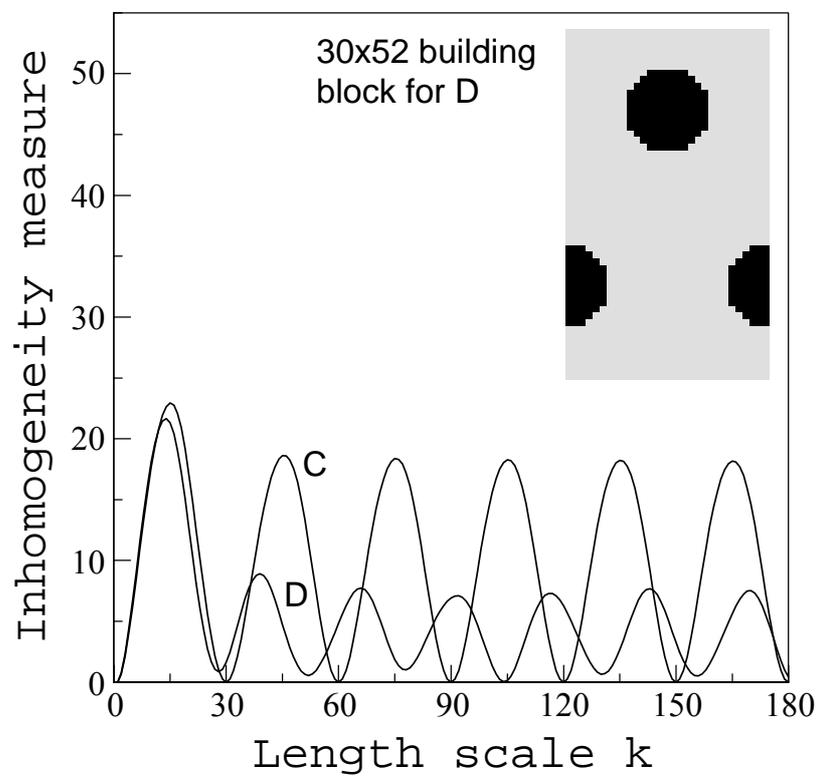

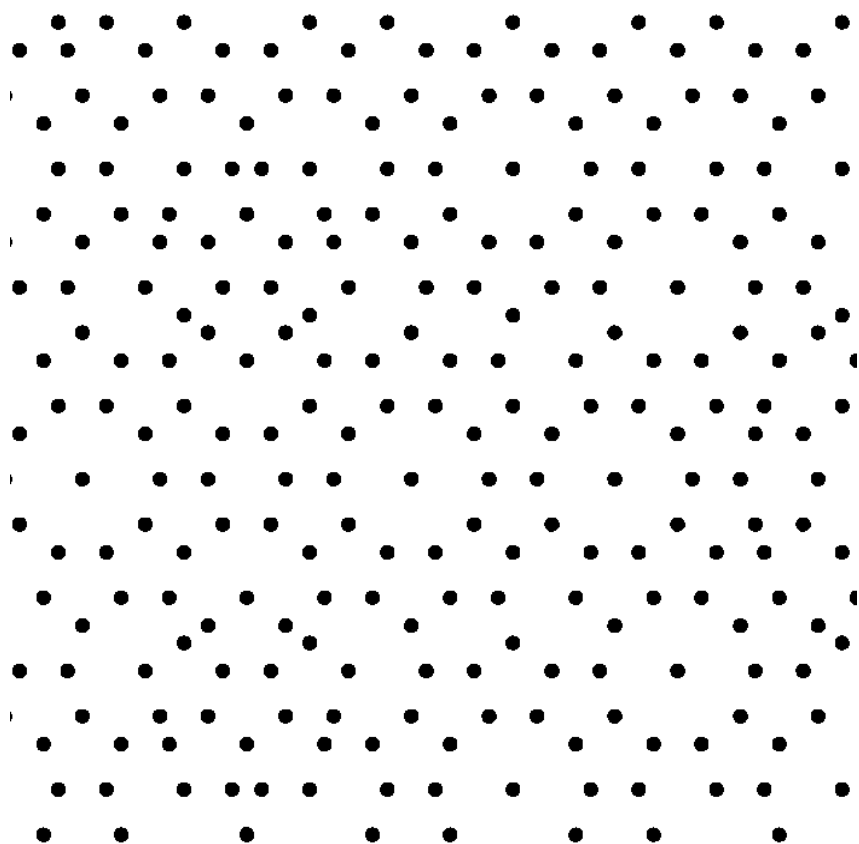

Fig. 4a

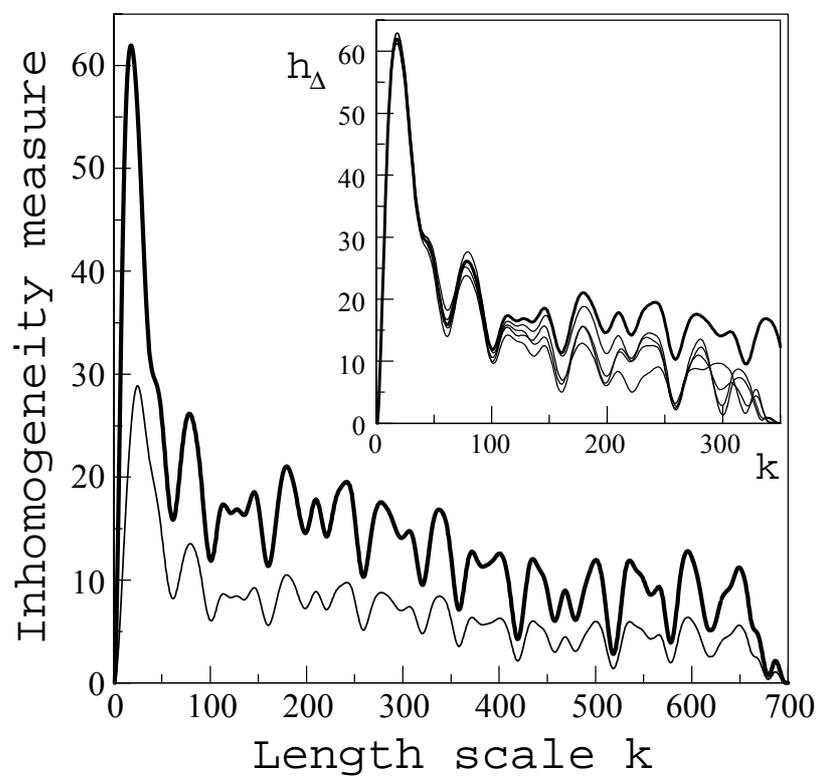

Fig. 4b

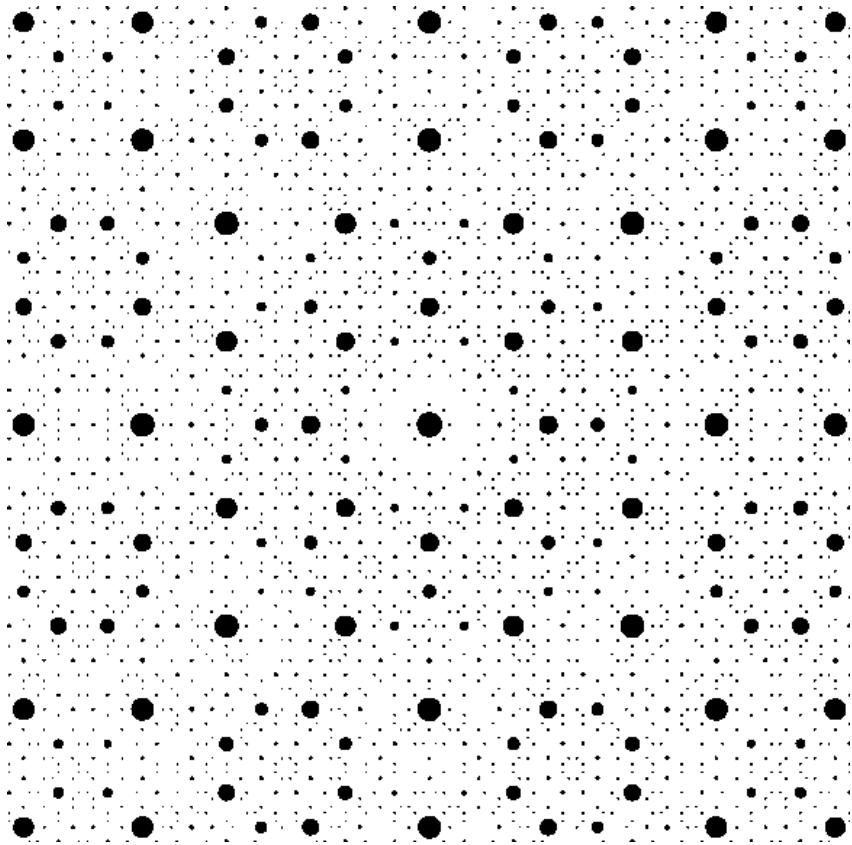

Fig. 5a

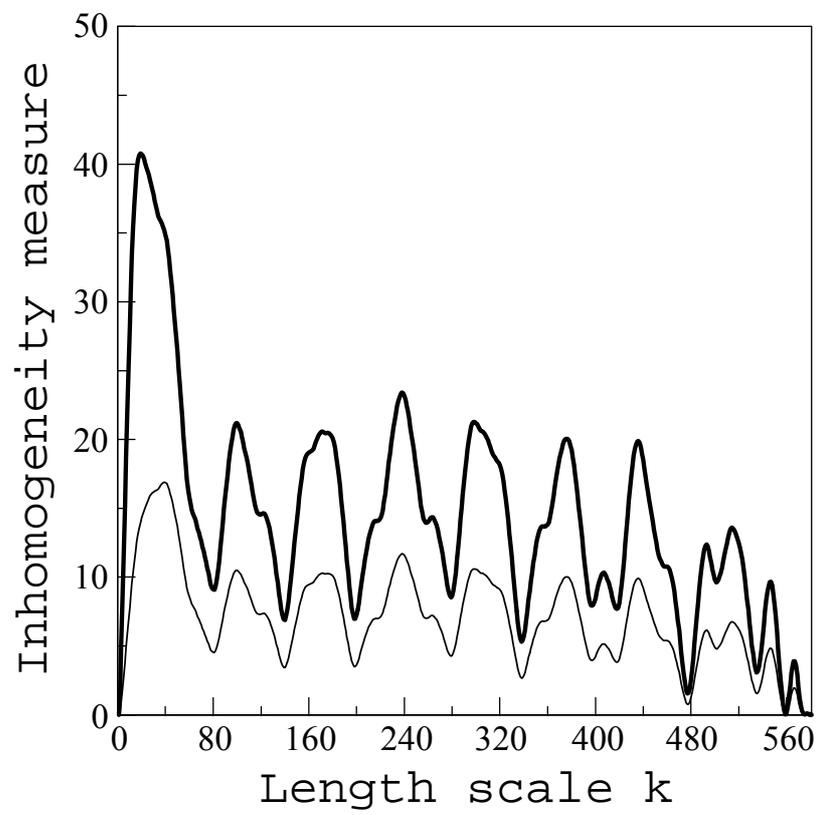

Fig. 5b